\documentclass{article}
\usepackage{amssymb}
\usepackage{amsmath}
\usepackage[utf8]{inputenc}
\usepackage{graphicx}
\usepackage{xcolor}
\usepackage{verbatim} %for comment block 

\usepackage{subcaption}

\newtheorem{theorem}{Theorem}

\newcommand{\etal}{\mbox{\emph{et al.\ }}}

\title{DCJVis: visualization of genome rearrangements using DCJ operations}

\begin{document}%\sloppy
\author{
Sruthi Chappidi\thanks{
University of Texas at Dallas,
800 W Campbell Rd, 
Richardson, Texas 75080.
\texttt{sruthi.chappidi@utdallas.edu, besp@utdallas.edu}}
\and
Sergey Bereg$^*$
}
\maketitle
\begin{abstract}
The {\em double-cut-and-join} (DCJ) operation, introduced by Yancopoulos \emph{et al.}, allows minimum edit distance to be computed by modeling all possible classical rearrangement operations, such as inversions, fusions, fissions, translocations and transpositions, in linear-time between two genomes. However, there is lack of visualization tool that can effectively present DCJ operations that will help biologists to use DCJ operation.  In this paper, a new visualization program is introduced, DCJVis, to create a diagram of each DCJ operation necessary to transform between the genomes of two distinct organisms by describing a possible sequence of genome graphs based on the selected gene adjacency on the source genome for the DCJ operation. 
Our program is the first visualization tool for DCJ operations using circular layout. 
Specifically, the genomes of \textit{Saccharomyces cerevisiae} and \textit{Candida albicans} are used to demonstrate the functionality of this program and provide an example of the type of problem this program can solve for biologists.

\end{abstract}

%\keywords{Comparative Genomics; DCJ operation; genome rearrangements, genome graphs }

\section{Introduction}

Comparative genomics compares the genome sequences of various species and reveals regions of similarity and dissimilarity based on the unique characteristics of the organism \cite{Lin,Periwa}. Through this analysis, conserved genes or synteny blocks of lineage species like {\em Saccharomyces cerevisiae} and {\em Candida albicans} can be discovered. Genomic rearrangement operations include inversions, transpositions, circularizations, linearizations, translocations, fusions, and fissions. Yancopoulos \etal \cite{yaf05} introduced a universal operation called the {\em double-cut-and-join} (DCJ) which generalizes these operations. This model supports genome rearrangements of single and multichromosomal genomes with circular or linear chromosomes, and provides a simple algorithm for computing the genomic distance between two genomes using a minimum number of DCJ operations. This minimum number helps in answering the evolutionary relationship between organisms.  
Bergeron \etal \cite{bms06} introduced an {\em adjacency graph} that simplifies the theory and distance computation considerably with the genome represented as a set of adjacencies and telomeres, or end points. 

%\cite{daSilva2013,miklos10}

The incorporation of insertions and deletions of chromosomes and chromosomal intervals (collectively called {\em indels}) into DCJ distance was 
was studied in \cite{braga10,Compeau2013,yf09}.
Yancopoulos and Friedberg \cite{yf09} extended the double-cut-and-join  operation paradigm to include genome rearrangements of partially matching genomes. They introduced a method for genomes with genes which are completely or partially unmatched in the other.
"Ghost adjacencies" are used to supply the missing gene ends in the genome not containing them. 
The method generalizes DCJ operations on the generalized adjacency graph, and provide a prescription for calculating the DCJ distance for the expanded repertoire of operations, which includes insertions, deletions, and duplications. 
Braga \etal \cite{braga10} developed a linear time algorithm to compute the genomic distance, considering DCJ
and indel operations. They found a preliminary evidence of
the occurrence of clusters of deletions in the {\em Rickettsia} bacterium.
Compeau \cite{Compeau2013} found a simple presentation of DCJ-indel sorting that still yields
a linear-time solution to the problem.

Seoighe \etal \cite{seoighe00}
%\In \cite{seoighe00}, the authors 
describe small inversions that are frequent in eukaryotes and cause drifting in the position and orientation of genes during speciation.The comparison study of the gene order evolution of the yeast species {\em S. cerevisiae} and {\em C. albicans} reveals inversion as the major rearrangement, leading to changes in gene location while still conserving sequence. The longest syntenic region, a region of significant similarity between compared organisms, is between Chromosome 3 in {\em S. cerevisiae} and Chromosome 7 in {\em C. albicans}, with high gene conservation revealed by gene mapping of both organisms, and this example will be used to explain how DCJVis approaches visualizing a DCJ solution \cite{chibana05}. Even though a conserved linkage between genes of {\em C. albicans} and {\em S. cerevisiae} exists, it is known to take a minimum of 4 inversions and 3 indels (insertions or deletions) to complete the genome rearrangement between the organisms. 

This paper demonstrates the ability of the novel software, DCJVis, to visualize the gene synteny and rearrangements between the closely related genomes of {\em S. cerevisiae} and {\em C. albicans} using the double-cut-join model introduced by Yancopoulous et al. [5]. DCJVis is an interactive circular genome viewer implemented in Java. This tool provides a clear presentation of each individual operation to the experimenter needing to use this program. As the DCJ model represents all the basic rearrangement operations, including inversions, fusions, and fissions on a uni-chromosomal genome, this gives biologists the option to visualize the operations and facilitates more efficient research approaches to large scale genetic rearrangement problems. There is extensive research as shown in  \cite{bms06,braga09,friedberg08,yf09,Shao15,Shao12,yaf05}papers with emphasis on theory aspect of DCJ operation, but not much focus on application development that will be useful for researchers in the area of comparative genomics. The other software packages and existing visualization applications in Table \ref{Tools_Comparison} currently do not provide all the functionalities of DCJVis. Our application, DCJVis comes with few unique features compared to existing DCJ operation visualization tools such as the gene representation using distinct colors, circular genome representation layout, interactive application with tool-tip feature to provide information of each gene, and accepts fasta input format provided by NCBI databases. 

\section{DCJ operation on linear chromosome and graphs}

\begin{figure}
\centering
\begin{subfigure}{.5\textwidth}
  \centering
  \includegraphics[width=.98\linewidth]{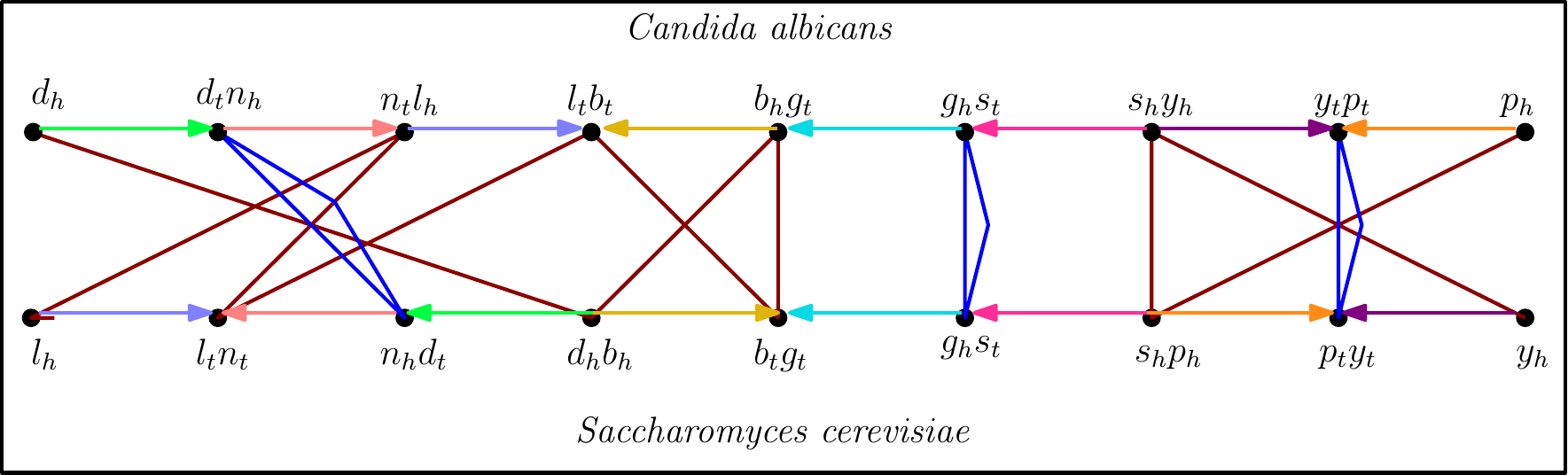}
\caption{ }
\label{AdjacencyGraph}
\end{subfigure}%
\begin{subfigure}{.4\textwidth}
 \centering
 \includegraphics[width=.5\linewidth]{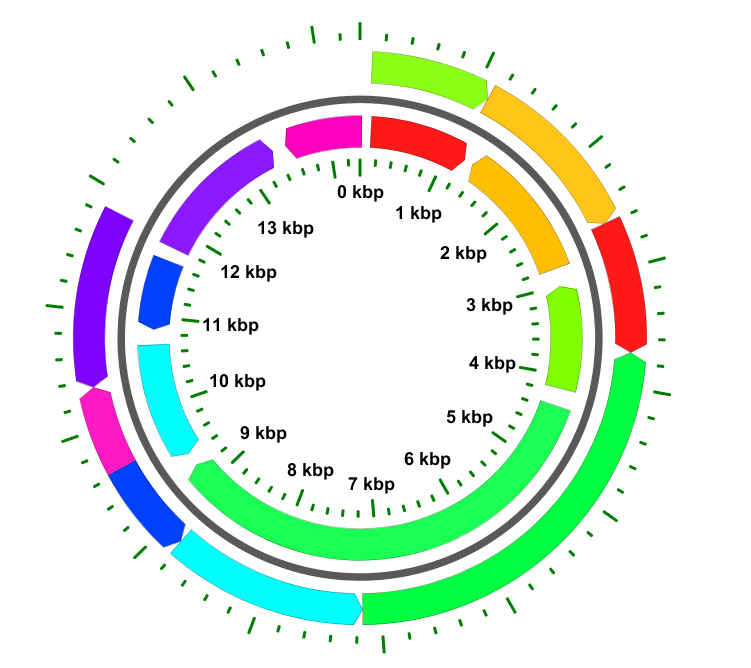}
\caption{}
\label{DCJVis_AG}
\end{subfigure}%
\caption{(a) \label{fig: main graph}Adjacency graph of $Candida$ $albicans$ and $Saccharomyces$ $cerevisiae$. (b) DCJVis alignment of the Adjacency Graph showing the target as the inner color wheel and the source as the outer color wheel circle, where each vertex represents a gene using a distinct color.}
\end{figure}

%used above as 2 sub images to align side-by-side
\begin{comment}
 \begin{figure*}
\centering
\includegraphics[scale=0.8]{main_Adjacency_graph}
\caption{\label{fig: main graph}Adjacency graph of $Candida$ $albicans$ and $Saccharomyces$ $cerevisiae$ }
label{AdjacencyGraph}
end{figure*}

\begin{figure*}
\centering
 \includegraphics[scale=0.7]{pic1_step0.png}
%\includegraphics[height=\textheight, width=\textwidth,keepaspectratio]{Idea1_v2.pdf}
\caption{A visualization tool alignment of the Adjacency Graph showing the target as the inner color wheel and the source as the outer color wheel circle, where each vertex represents a gene using a distinct color.}
\label{DCJVis_AG}
\end{figure*}
   
\end{comment}

\subsection{Background}
The terminology of Bergeron \etal \cite{bms06}, putative biology terms, are used to describe the genome notation shown on the graphs with genes listed as vertices.
The genome consists of many Open Reading Frames (ORFs), which are long sequences of DNA with coding regions resulting in a gene. A series of genes linked by adjacencies form a chromosome, and one or more chromosomes form the entire genome of an organism. 

For simplicity, ORF synteny blocks of {\em S. cerevisiae} and {\em C. albicans} \cite{chibana05} are represented as genes denoted with single letter $a$. 
Sequences of DNA with tail and head extremities are used as boundaries for the genes. 
The head and tail extremities of a gene $a$ are denoted by $a_h$ and $a_t$, respectively\footnotetext{The tail refers to the 3' end of a gene and the head refers to the 5' end in biology.}. 
An {\em adjacency} of two consecutive genes $a$ and $b$ is the pair of their adjacent extremities.  
An extremity of a gene $a$ that is not adjacent to any other gene is called a {\em telomere}, denoted by $\{a_h\}$ or $\{a_t\}$, see also \cite{bms06,friedberg08,kothari07}. 
Graphs with a vertex of degree one are used to represent telomeres and graphs with a vertex degree of two represent adjacencies. 
Each DCJ operation results in a change to some part of the genome graph. 
If genes $a$ and $b$ are adjacent, there can be four different types of their adjacency depending on the orientation of the genes $a$ and $b$: $$\{a_h,b_t\},\{a_h,b_h\},\{a_t,b_t\},\{a_t,b_h\}.$$

\subsubsection{Genome Graphs}
A genome, as defined in this program, is a set of adjacencies and telomeres such that the tail or the head of any gene appears in exactly one adjacency or telomere \cite{bms06,friedberg08}. 
A {\em genome graph} can be constructed as follows: the telomeres and adjacencies of the genes are the vertices of the genome graph. Edges in the genome graph are used to represent genes, and connect two vertices corresponding to the tail and the head of the gene. The connected components of the genome graph correspond to {\em chromosomes}. 
This characterization of genetic information allows for a visualization of the structure and orientation of genes in both linear and circular chromosomes where each vertex is represented as an adjacency or telomere with the head and tail of a gene. The chromosomes of {\em S. cerevisiae} and {\em C. albicans} will be used to visualize the DCJ operation, and will be bounded by telomeres\cite{friedberg08}. In Figure \ref{option1}, genome graphs are used depict the changes in vertices after each DCJ operation from step 1 through 4.

\subsubsection{Adjacency Graphs}
The {\em adjacency graph} of two genomes Genome1 and Genome2, denoted \textit{AG(Genome1,Genome2)}, is a bipartite graph with vertices corresponding to the extremities of the two genomes and edges connecting extremities with the same labels. 
The adjacency graph was introduced Bergeron \etal \cite{bms06}. 
It allows the user to visualize potential arrangements of the genes of a genome. 
Connecting the adjacencies or telomeres between two similar genomes will result in connected paths and cycles with linear or circular chromosomes of the genomes. 
The adjacency graph of two genomes can be computed using Algorithm 1 proposed by Bergeron \etal \cite{bms06} by simply joining the endpoints of each gene from the initial genome to the target genome starting with an adjacency mapping and followed by telomeres.
For example, the adjacency graph of the {\em C. albicans} and {\em S. cerevisiae} genomes is depicted in Figure \ref{AdjacencyGraph}. 
Any of the adjacencies depicted on an adjacency graph that differ between the source and target genomes can be used as a starting point to find a possible sequence of sorting using the DCJ algorithm. 
The graph also shows the position of each gene between the selected start and end genomes \cite{friedberg08}. 

To perform DCJ operations for a genome rearrangement between any two genomes, the set of genes within each of the two compared genomes must be the same. 
Adjacency graphs resulting from a union of cycles and odd paths between the two genome graphs list the steps necessary to transform Genome1 to Genome2 by DCJ operations. 
For example, the two genomes are equal if and only if $N = C+I/2$, where $C$ is the number of cycles, $N$ is the total number of genes in each genome, and $I$ is the number of odd paths in the adjacency graph \cite[Lemma 1]{bms06}.

\subsection{DCJ Operation}
\begin{figure}[htb]
\centering
\framebox{\includegraphics[scale=0.55]{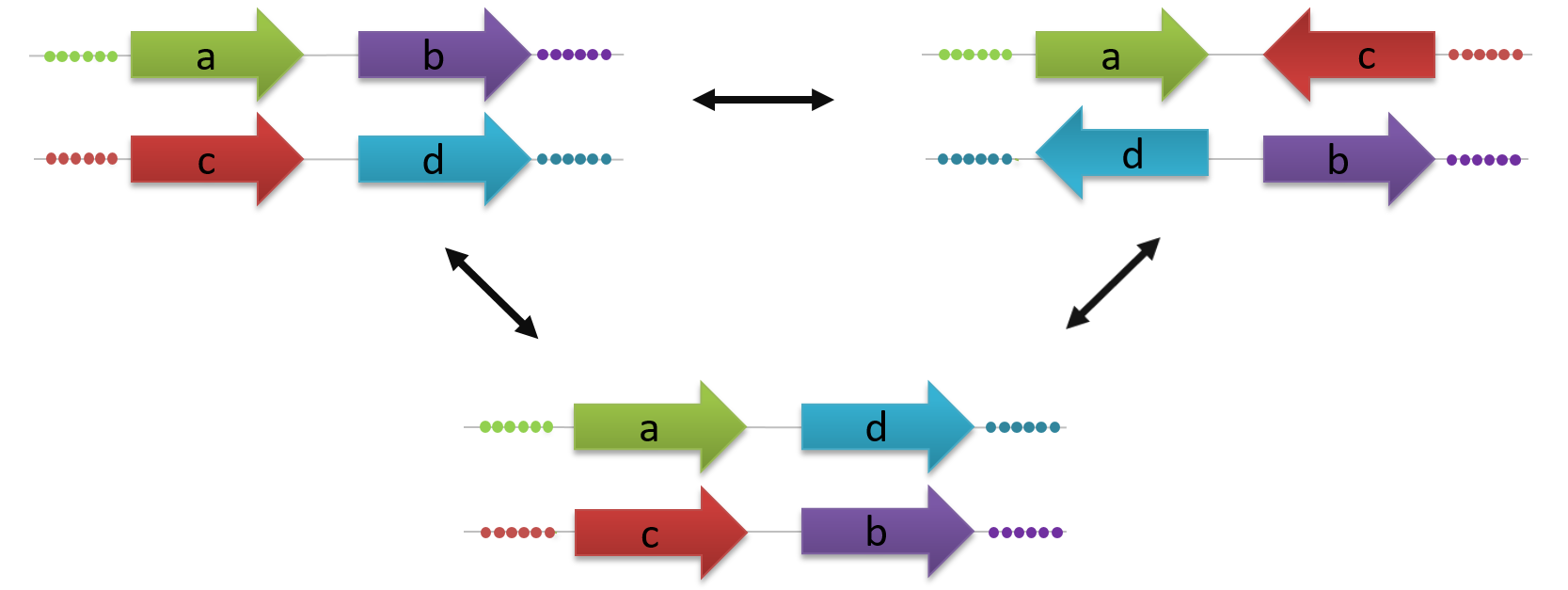}}
\caption{General DCJ operation with linear chromosomes. The orientation of a gene can change but the orientation change should propagate for the adjacent genes}
\label{fig:DCJ}
\end{figure}
The DCJ model, as it can describe all the genomic rearrangements, has formed the basis for most algorithmic research in the area of Comparative Genomics \cite{bms06,yaf05}. 
A simple linear-time algorithm for computing DCJ distance between two genomes, it calculates the minimum of number of genome rearrangement operations required to transform one genome into another. 
A DCJ operation consists of cutting two vertices in the first genome, as demonstrated in genome graph in  Figure \ref{option1}, producing four cut ends, then rejoining the four unconnected vertexes in a different order to produce the desired vertices of the target genome (Figure 1). 
As described in \cite{braga09,yaf05}, there are four possibilities for a DCJ operation on an adjacency or telomere: 
\begin{enumerate}
\item A DCJ operation on adjacencies $ab$ and $cd$ would result in new adjacencies such as $ac$ and $bd$ or $ad$ and $bc$, which defines an {\em inversion} operation on a linear chromosome or a {\em translocation} operation on a genome with two linear chromosomes. See Figure \ref{fig:DCJ} and \ref{fig:dcj1}. In Figure \ref{fig:DCJ}, the orientation of a gene can be different than depicted, but the change might have effect on the adjacent genes orientation.  
\begin{figure*}[htb]
\centering
\includegraphics[scale=0.6]{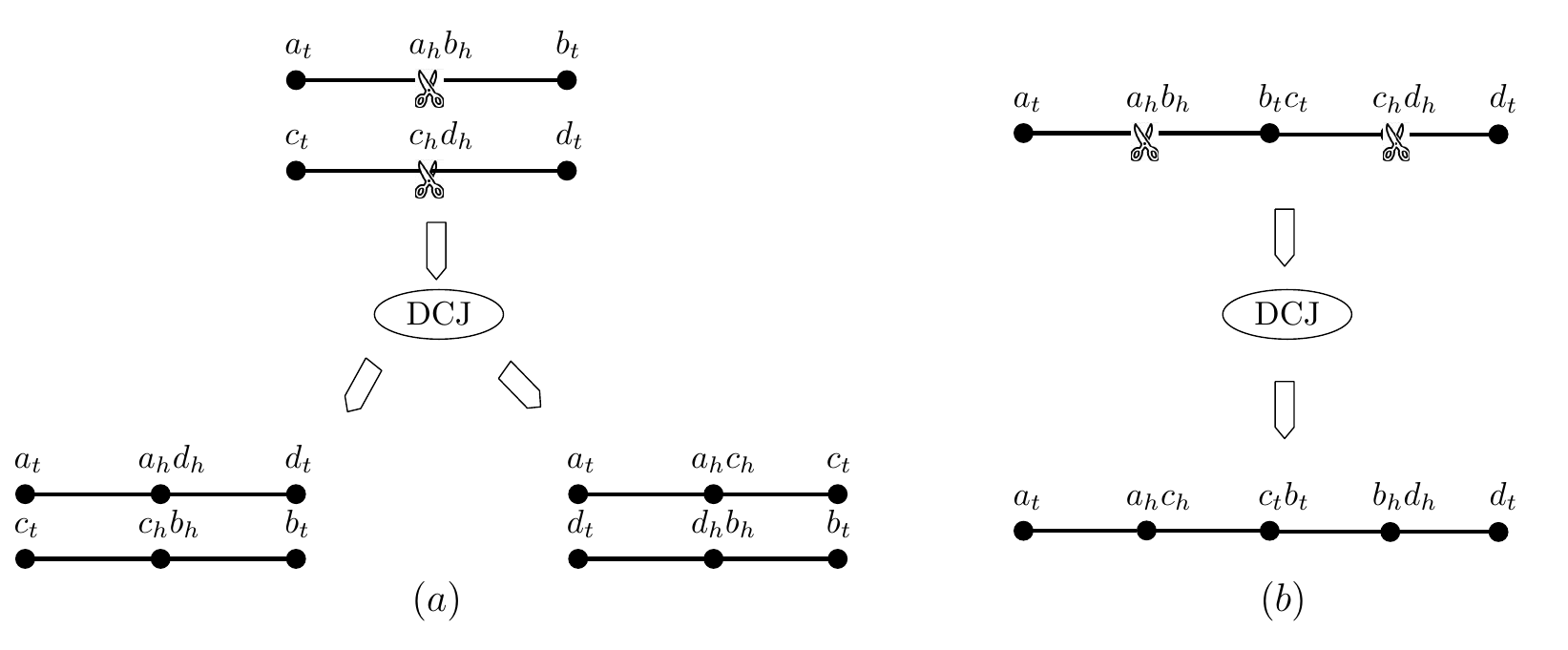}
\caption{(a) {\em Translocation} operation yielded by a DCJ operation on a genome with two linear chromosomes. (b) The DCJ operation yields an {\em inversion} operation on uni-chromosomal genomes.}
\label{fig:dcj1}
\end{figure*}

\item  A DCJ operation on a single adjacency $ab$ and a telomere $c$ would create either $ac$ and $b$, or $a$ and $bc$, which correspond to an {\em inversion} or a {\em translocation} (including {\em fission} and {\em fusion} of circular and linear chromosomes).
\begin{figure}[htb]
\centering
\includegraphics[scale =0.6]{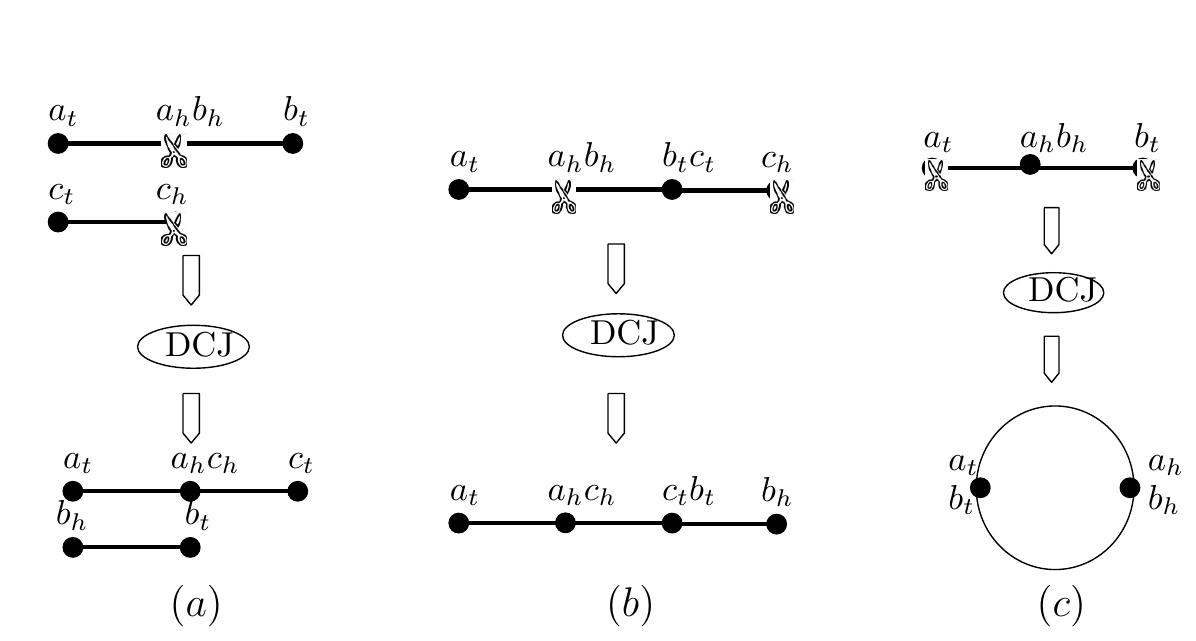}
\caption{(a) A $translocation$ operation on a genome with two linear chromosomes, (b) an $inversion$ operation on a linear chromosome with $a,b,c$ genes and (c) $Circularization$ of a genome with a linear chromosome.}
\label{fig:dcj2}
\end{figure}

\begin{comment}
\item  A DCJ operation on two telomeres, $a$ and $b$, results in a new adjacency $ab$, which is a {\em fusion} operation on two linear chromosomes resulting in circularization.
\begin{figure}[htb]
\centering 
\includegraphics[scale=0.8]{dcj3}
\caption{$Circularization$ of a genome with a linear chromosome.\textbf{try to combine the figure 5 and 6 so }}
\label{fig:dcj3}
\end{figure}
\end{comment}
\item A DCJ operation on only one adjacency $ab$ can result in the creation of two telomeres $a$ and $b$ reflecting a {\em fission} operation on a linear chromosome resulting in two chromosomes or on a circular chromosome resulting in linearization.
\end{enumerate} 
\begin{figure}[htb]
\centering
\includegraphics[scale=0.6]{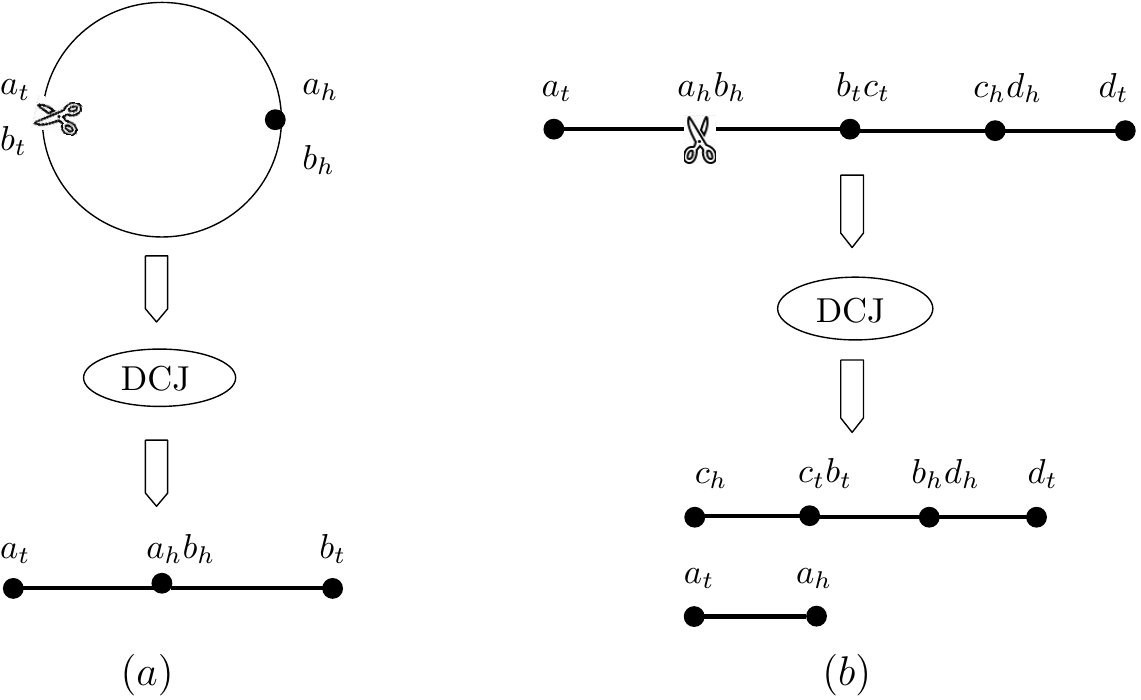}
\caption{(a) $Linearization$ and (b) $fission$ operations yielded by DCJ.}
\label{fig:dcj4}
\end{figure}

Given two genomes $G_1$ and $G_2$ defined on the same set of $N$ genes,  the length of the shortest sequence of DCJ operations that transforms  $G_1$ into $G_2$ is called the {\em DCJ distance} between $G_1$ into $G_2$, denoted by $d_{DCJ}(G_1,G_2)$. The {\em consistent adjacency graph} of the two genomes $G_1$ and $G_2$, see Figure \ref{AdjacencyGraph} for an example of two genomes. The genomes {\em C. albicans} to {\em S. cerevisiae} will be used to find one of the possible optimal DCJ sorting sequences. Bergeron \etal \cite{bms06} proposed a greedy sorting algorithm for computing the DCJ distance in linear time based on the following result.

\begin{theorem}[Bergeron \etal \cite{bms06}]
The DCJ distance between two genomes $G_1$ and $G_2$ defined on the same set of $N$ genes can be computed as 
\begin{equation}\label{eq:dcj}
 d_{DCJ}(G_1,G_2)= N-(C+ I/2),
\end{equation}
where $C$ is the number of cycles and $I$ is the number of odd paths in the adjacency graph.
\end{theorem}
Four DCJ operations are required for the complete genome rearrangement from {\em C. albicans} to {\em S. cerevisiae}, based on two odd paths, three cycles, and eight genes as shown in Figure \ref{fig: main graph}.

\section{Biological Purpose}
The physical location of a gene within the context of the genome can affect the regulation of that gene. Rearranging synteny elements of chromosomes between one organism and another is one of the ways biologists can study how the regulation of homologous genes across organisms contributes to the unique identity of an organism. Additionally, understanding how the position and orientation of each gene affects expression allows scientists to create synthetic organisms with novel arrangements of genes which can efficiently perform a desired function. \cite{Lin} A convenient way to visualize each step of a DCJ operation would facilitate projects like these.

The visualization of this tool needs to provide easy and rapid understanding of large amounts of data, such as the number of rearrangements needed to transform one genome into another, the operation used in each step, and a clear picture of what has happened between each step in the process.

The new algorithmic DCJ model, which is simple in nature, was introduced to find optimal sorting sequence in O(N). When compared to the inversion operation that was used for decades in research on comparative genomics, the DCJ model proved to be a simpler and faster alternative \cite{chibana05}.

\begin{figure*}[ht]
\centering
\includegraphics[scale=0.55]{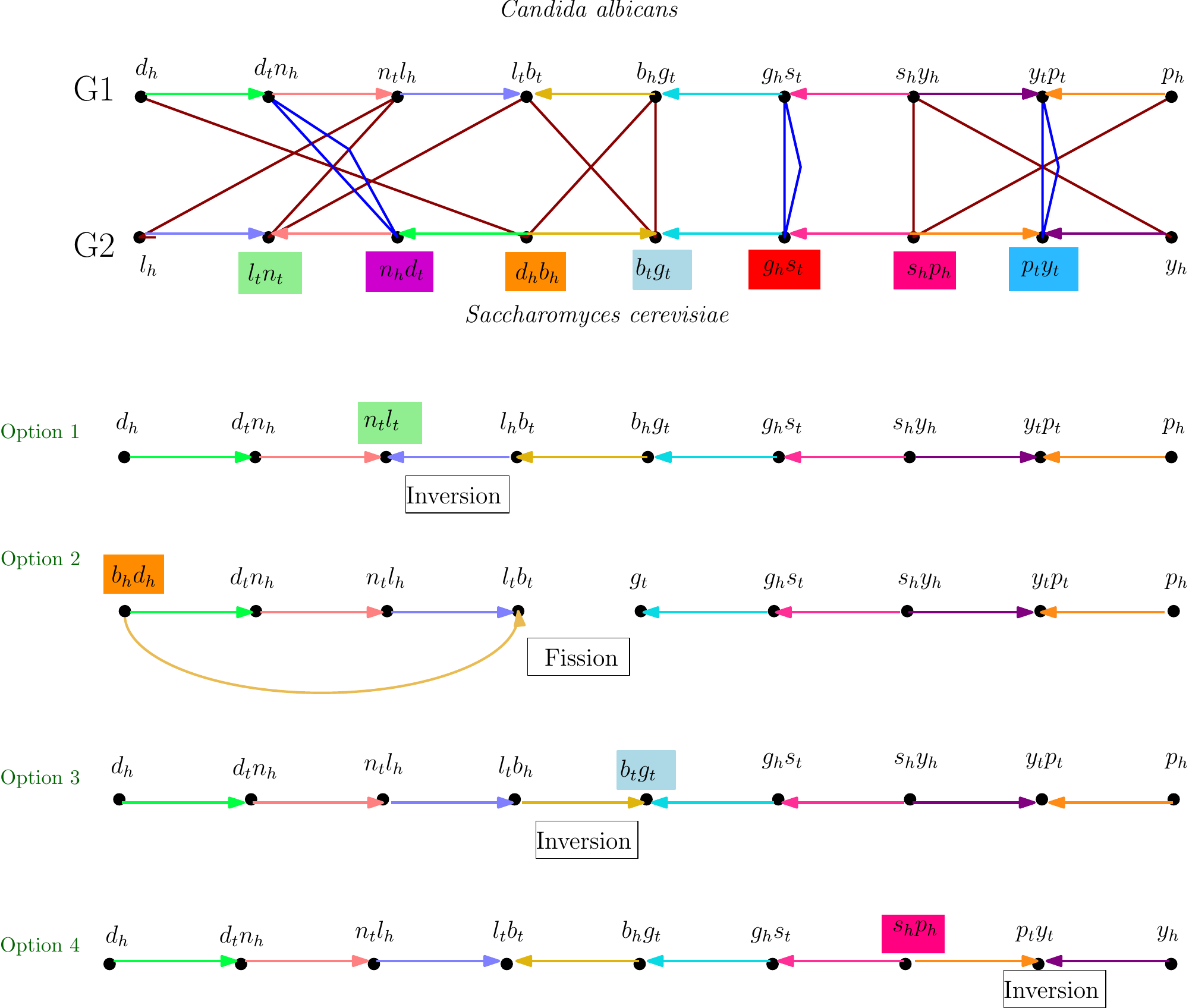}
\caption{First DCJ operation with various adjacency vertex as starting vertex in transforming genome of {\em Candida albicans} into {\em Saccharomyces cerevisiae}. The adjacency vertices $n_td_h$, $g_ts_h$, and $p_hy_h$ are in same orientation in both genomes. As a result, there are only four adjacencies that are different between both genome graphs to be considered as possible options for first DCJ operation, resulting in four options for the first DCJ operation.}
\label{Options_IPE}
\end{figure*}

\begin{figure*}
\centering
\includegraphics[scale=0.55]{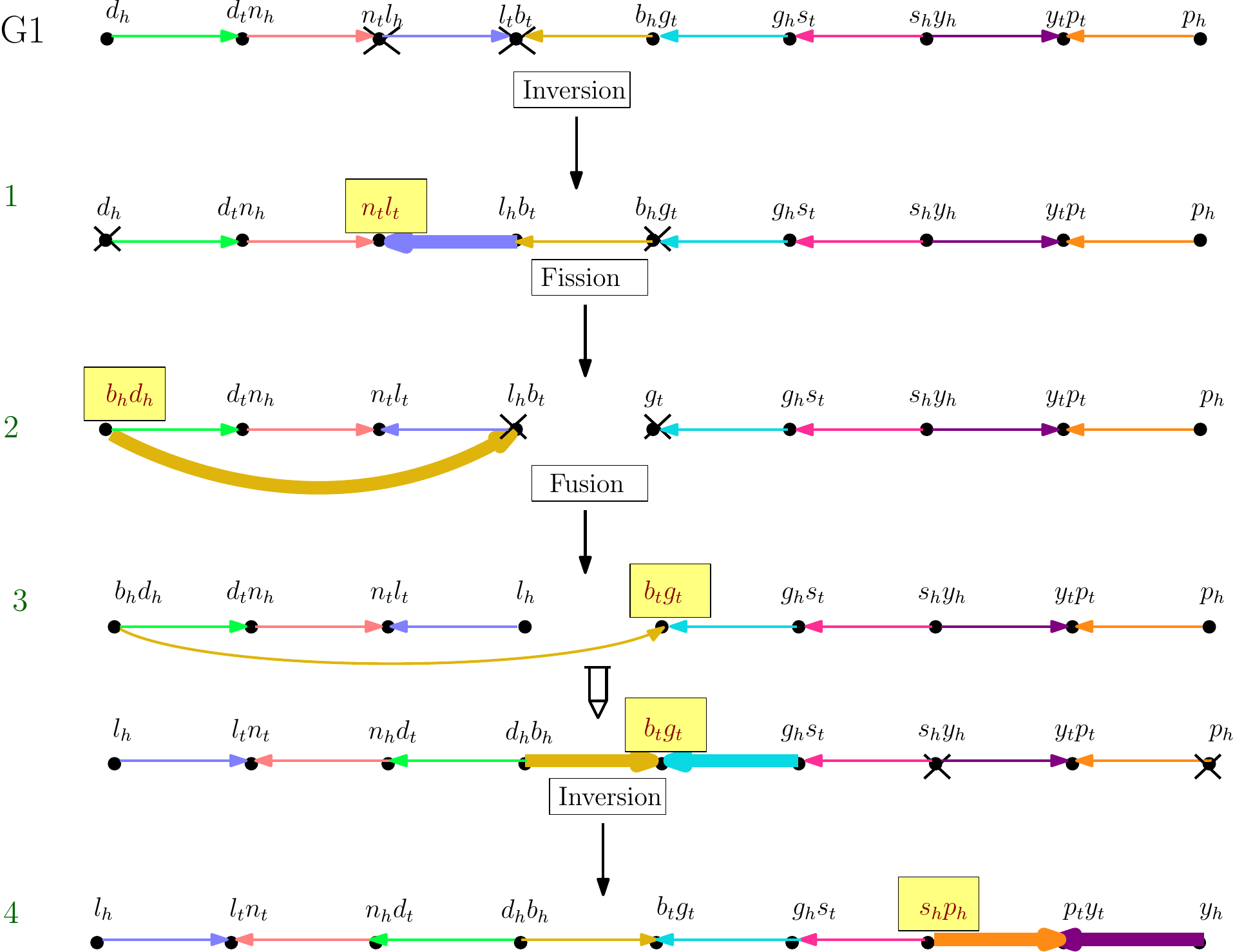}
\caption{Scenario 1 $-$ Complete set of genome rearrangement operations to transform {\em C. albicans} into {\em S. cerevisiae} started from target adjacency as in option 1 of figure \ref{Options_IPE} with {\em inversion, fission, fusion and inversion}.}
\label{option1}
\end{figure*}

Using the DCJ operation algorithm \cite{bms06}, the current way to represent the information given by the algorithm is to either hand-draw a representation, or to use a basic visualization program, such as Ipe software editor, to show a diagram of each step. 
Figure \ref{Options_IPE} shows the Ipe representation of the possible initial steps of the DCJ algorithm solution for the example problem, the rearrangement of \textit{Candida albicans} chromosome 7 to \textit{Saccharomyces cerevisiae} chromosome 3. 
This visualization allows the scientist to see each step more clearly than the algorithm output, however each step needs to be manually drawn to get this representation. The UniMoG software \cite{UniMoG} allows the user to input a text file of the initial genome and the desired end product genome, and will then create a visual representation of each step necessary to convert between the two. 
Colors are used by this program to highlight the regions undergoing a change in each step, but the colors used do not have meaning and are not representative of any specific segment of the genome. 
The main difficulty in using this program is that the genome input needs to be formatted to one of the accepted options. 
Genome files obtained from the standard online databases, such as NCBI, are not in one of the accepted formats and need to be manually converted by the user. 
The UniMoG program has been tested with genomes up to 32,500 genes, but the visualization of the solutions for complex problems is difficult, requiring the user to scroll through the linear depiction until the portion of the genome affected in that step is found. 
This is main disadvantage as the user will not be able clear identify the region of change efficiently. 
DCJVis was developed to help answer some of the difficulties encountered by previous visualization methods. 
When given two fasta file genomes to convert between, this program will process the raw fasta files by checking constraints, such as total count of genes, size of each, and position, into an acceptable format to utilize the DCJ distance calculation of UniMoG. 
UniMoG then finds the first DCJ solution scenario for the transformation, which DCJVis will represent as a series of circular graphs as shown in Figure \ref{option1} and Figure \ref{allsteps}. 
DCJVis provides complete overview of the entire genomes and point of change clearly visible for the user. 
Further, colors are used by this visualization process to designate each synteny region, allowing the user to easily spot the location of each synteny region across all steps \cite{Periwa}.

\section{Implementation}
\subsection{Color}
Each synteny region, or gene, is randomly designated a unique color by CGView. Once a color has been assigned to a gene, that color is maintained through all steps. The biologist can utilize this to track the position and orientation of genes throughout each of the steps. The changes in position and orientation of the genes which occur between each step are highlighted using focus to form a black outline around the altered genes (Figure \ref{Example_operations}).

\begin{figure*}
\centering
\framebox{\includegraphics[scale=0.6]{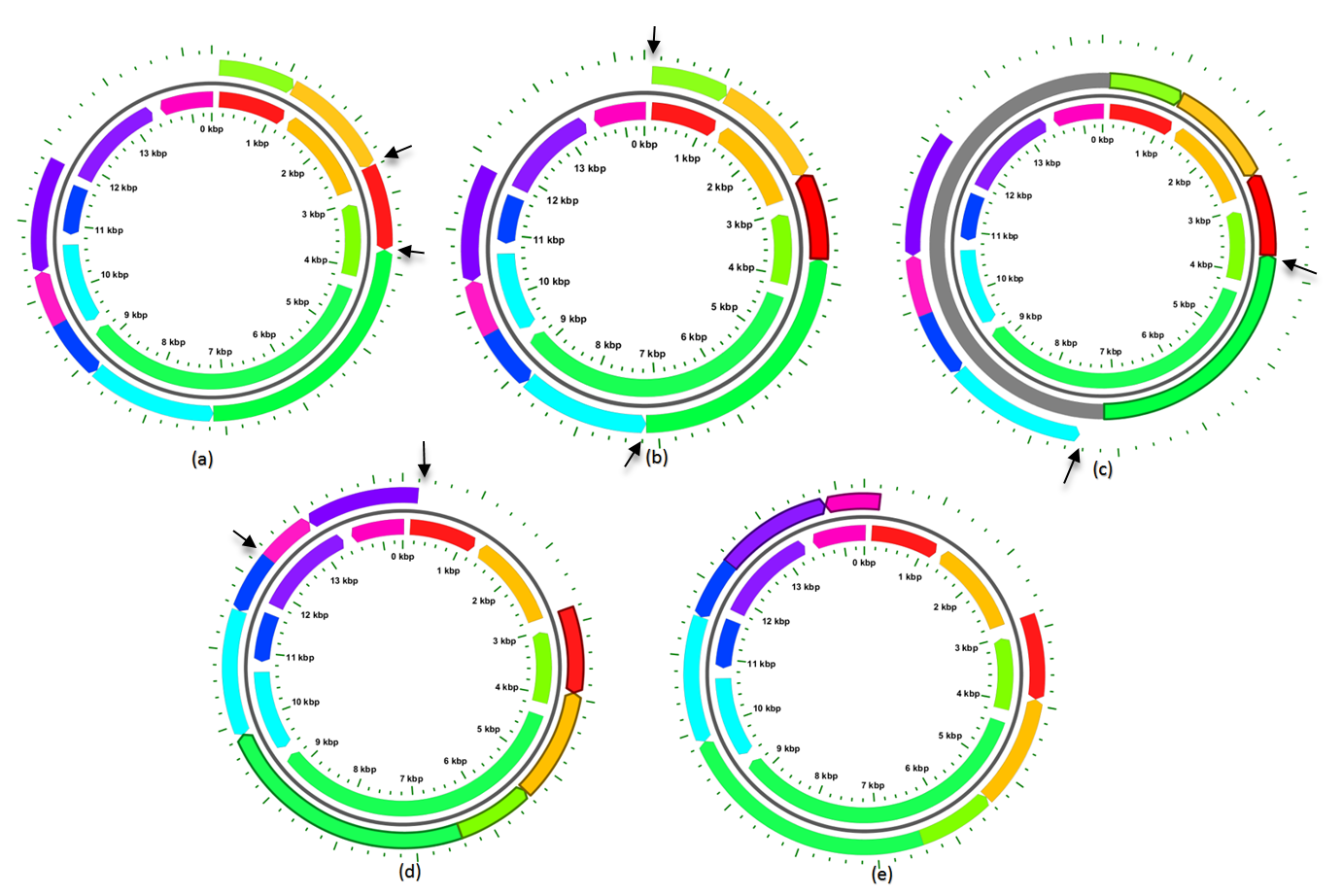}}
\caption{(a) Initial graph to show the rearrangement from {\em C. albicans} to {\em S. cerevisiae}. (b-e) four DCJ steps to complete transformation of {\em C. albicans} to {\em S. cerevisiae}.} 
\label{allsteps}
\end{figure*}

By assigning a unique color to each gene, the source and target genomes can be viewed as a puzzle where the pieces need to be rearranged in as few steps as possible. This visual representation conveys the discrepancies between each step in a concise way to the user. (Figure \ref{allsteps}).

\subsection{Data}
Multi-fasta file format is the universal format used by reviewed databases, such as NCBI, to record the nucleotide sequence of genetic information. DCJVis takes one multi-fasta for each genome as input to acquire the gene information along with the orientation and size, giving this program an advantage for biologists over the available alternatives. The existing UniMoG tool \cite{UniMoG} limits acceptable inputs to text files with genes listed either as alpha or numeric value, forcing biologists who want to use this program to manually convert the fasta files of their desired start and end genomes into an acceptable format. 

\subsection{DCJVis}
DCJVis is a desktop application developed in Java using the NetBeans Platform. DCJVis provides a helpful output that is readily understood using several pieces of internal architecture. DCJVis consists of a hierarchical zoom-able user interface which allows the user to zoom in, to get more detailed information, and zoom out for an overview. We use a "scene graph" model that is common to 3D environments, and CGview to maintain the hierarchical structure of objects and cameras, allowing the application developer to orient, group, and manipulate objects, and to upload large genomes \cite{Bederson04}. The genome is structured as a list of chromosomes, each with a different number of genes.
Each of the Nodes has a shape realizer at the lowest level which is rendered on the screen.
Each gene is realized by an Arrow shape, with the direction of the arrow indicating gene orientation and the length and position representing gene length and position.

DCJVis uses CGView\cite{Stothard} for image rendering and configuration to support visualization of circular genomes \cite{Stothard}. Its primary purpose is to serve as a component pipeline for generating visual output consisting of gene feature information and rendering options. Also, it is used to display the output of sequence analysis programs in a circular context. 

A circular representation is commonly used in biology to represent chromosomes or genomes in a space-efficient manner.  A circular expression map enables easy identification of expressed regions, and visually comprehensible \cite{Overmars15,Sato03}. This allows a viewer to easily see the relationship between all components of the chromosome at once and as explained by Ekdahl \etal \cite{Ekdahl04} circular layout provides many advantages including to give global overview of genome features. The CGView library was used to store the gene information as a collection of features. The CGView API's were used in the Java application to create circular gene maps. The main class used by the code is CGView, which represents a circular map of a genome sequence. Different methods were used in CGView that define the genome maps as {\em FeatureSlot}, defining the UI slot where the different gene features are stored.{\em Feature} is the back-end data structure to store the gene information. Each gene feature from the CGView data structure is stored as a node in the graphics hierarchy, which is accessed to draw the intermediate genomes during the transformation.{\em Legend} provides different front-end structures, defining the shape of the gene \cite{Stothard}.
\begin{figure*}
\centering
\framebox{\includegraphics[scale=0.4]{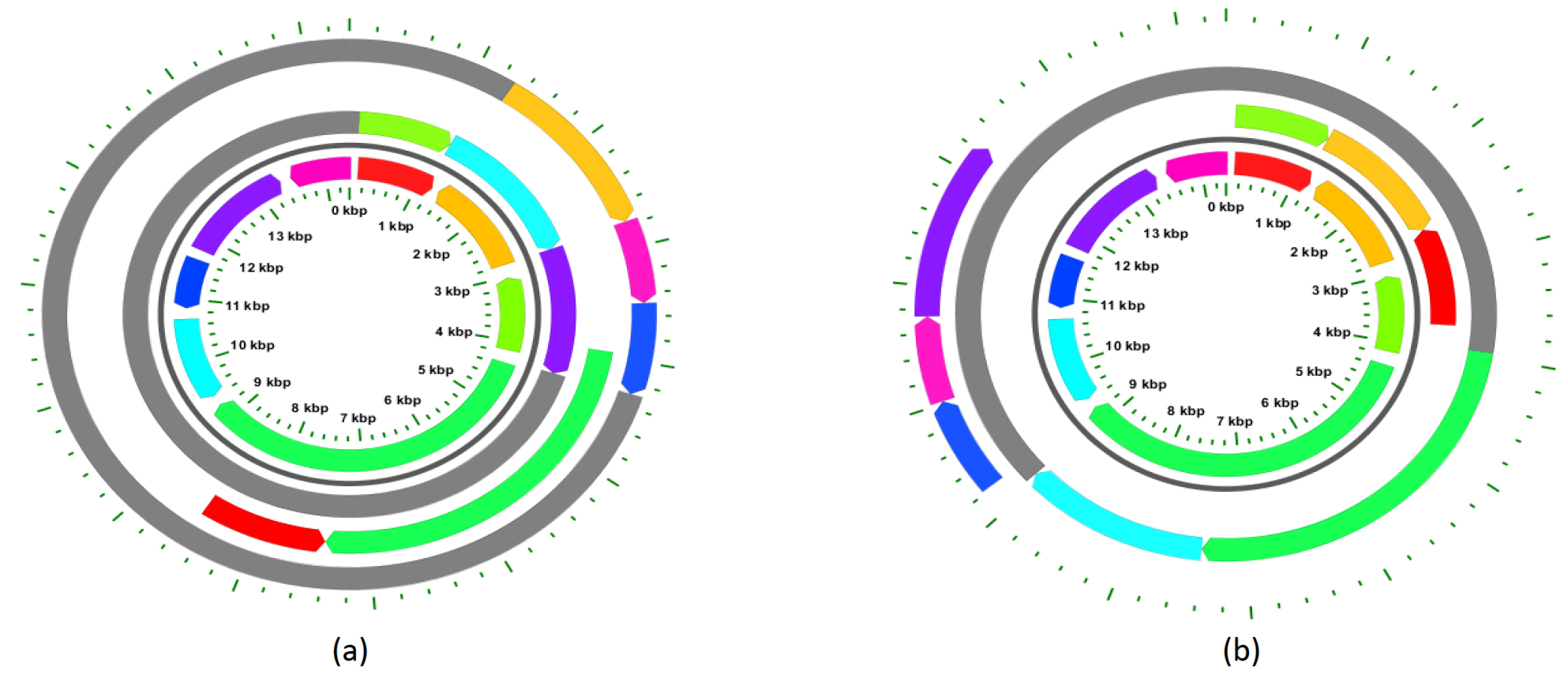}}
\caption{ Display of multiple chromosomes on increasing outer layers.}
\label{Multi_Chrs}
\end{figure*}
Step 1 : The parser for the Fasta File converts Fasta file data into one of the accepted formats and populates the internal data structure into UniMoG.
Step 2: The Map layer is defined as a collection of styles. 
Each layer has a Central backbone.
\begin{figure*}
\centering
\includegraphics[scale=0.5]{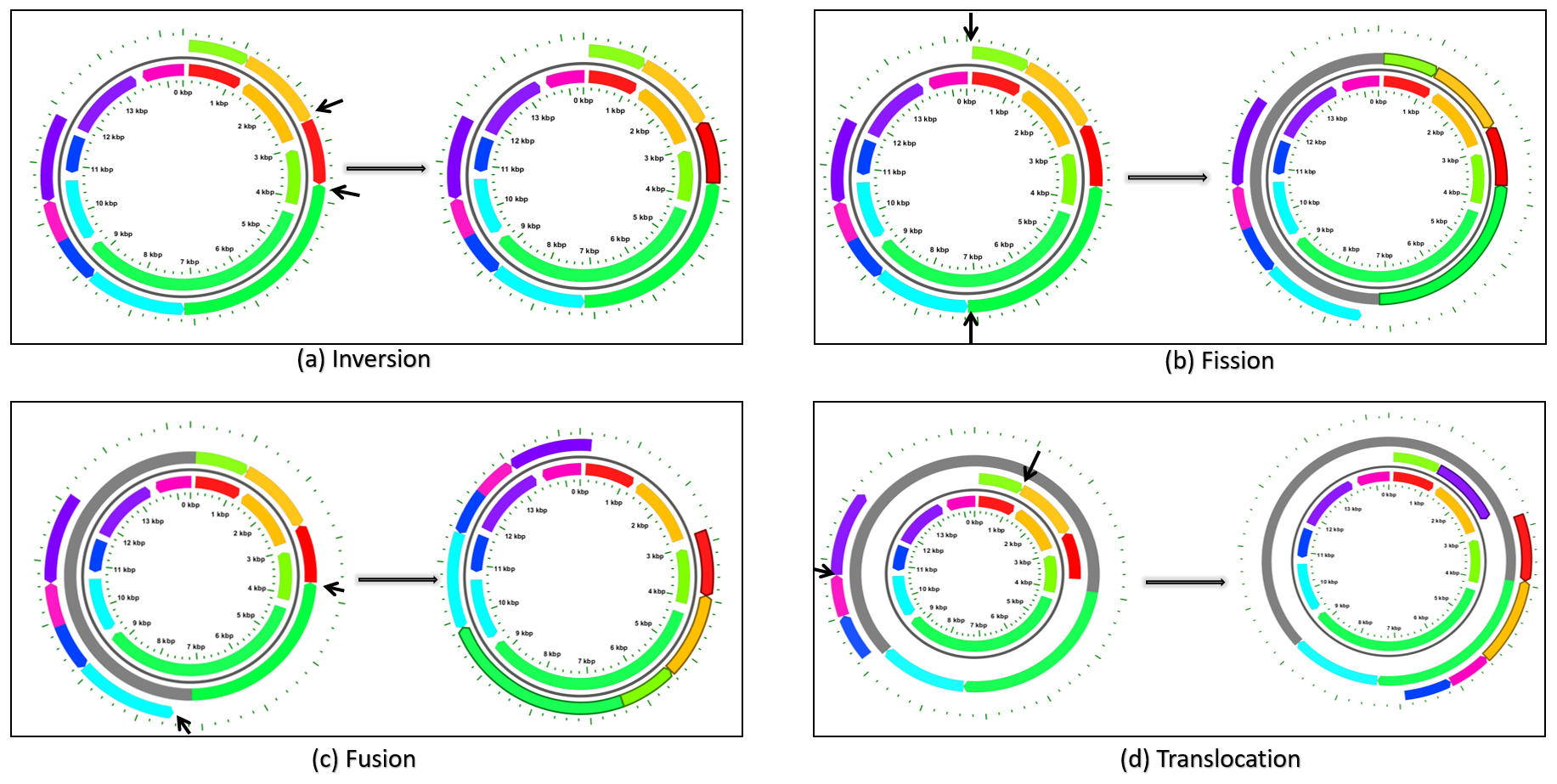}
  \caption{ Distinguishing the various genome rearrangement operations as result of DCJ operation: (a) Inversion, (b) Fission, (c) Fusion, (d) Translocation. The $\rightarrow$  used to clearly indicate location of double cut of the DCJ operation}
\label{Example_operations}
\end{figure*}
The target genome is represented as an inner circle with the source genome depicted as an outer circle. In steps resulting in multiple chromosomes, each new chromosome generated is shown as a circle in new outer layer (Figure \ref{Multi_Chrs}).
Each chromosome is defined by a genome slot.
Each gene is defined by a gene slot.
Step 3: The DCJ algorithm processes the input data of the two genomes provided.
Step 4: The output of DCJ algorithm  populates the output dataset, along with the intermediate DCJ steps.
Step 5: Each step is displayed on a separate frame.

\begin{figure*}
\centering
\framebox{\includegraphics[scale=0.5]{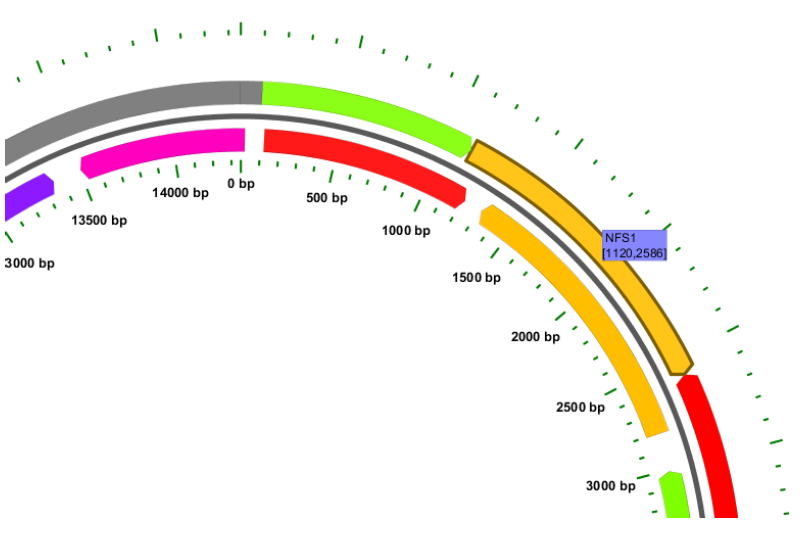}}
\caption{ Zoom-in for scalability and Tool tip for gene information}
\label{ToolTip_ZoomIn}
\end{figure*}

DCJVis enables the user to distinguish the operations performed at each step of DCJ operation leading to complete transformation of a source genome to a target genome as shown in Figure \ref{Example_operations}. An inversion operation is displayed by the inversion of the color arrow representing the segment that underwent inversion. A fission operation results in a new chromosome, represented by a new layer on top of the outer most layer. Fusion operations  are represented by decreasing the number of layers from the outer circle, or source genome. A translocation operation is shown by the movement of genes between layers, or chromosomes. In addition, this user interactive program, DCJVis, displays the updated gene locus and base pairs, and all the steps for transforming the source genome into the target genome as shown in Figure \ref{ToolTip_ZoomIn}. 

To best of our knowledge, there is no software visualization tool has been implemented that is interactive and adaptability to the large genome data based on DCJ algorithm proposed by Yancopoulos \etal \cite{yaf05}. 
As shown in Table 1, the comparison of features of DCJVis with existing visualization tool that are used widely for the genomic data visualization, such as CIRCO which used as standard for creating publication quality images, shows DCJVis to be unique tool with circular layout to illustrate the each step of genome rearrangement operations calculated based upon DCJ model to transform one genome into another. DCJVis demo may be accessed from http://utdallas.edu/~sxc105920/DCJVis/home.html. 

\begin{table*}
\centering
\framebox{\includegraphics[scale=0.6]{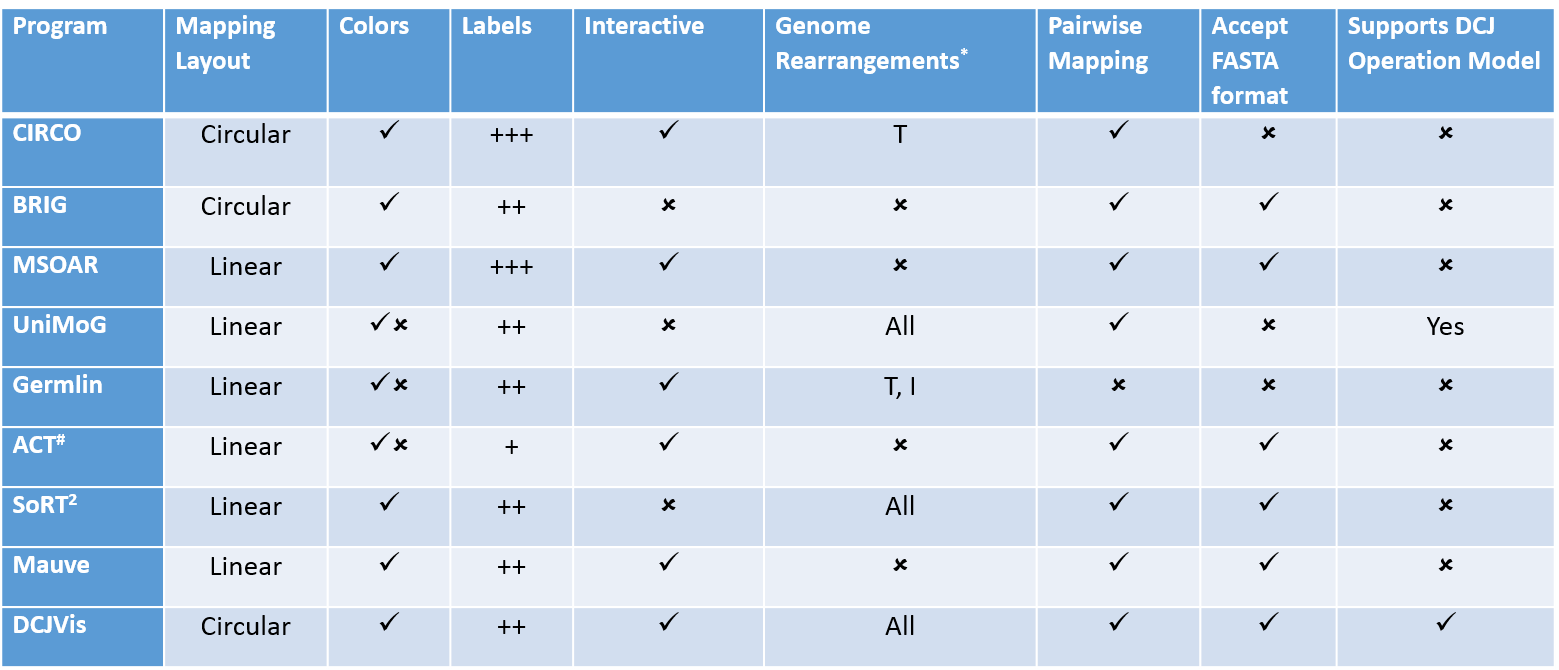}}
\caption{ Comparison of the features of DCJVis with existing genome visualization tools.}
\label{Tools_Comparison}
\end{table*}

\section{Conclusions}
DCJVis, a new visualization approach for the DCJ model genome rearrangement, allows the user to analyze the genome structure and the types of the DCJ operations. It can be used to find the optimal number of operations for computing the sequence of DCJ operations between two genomes \cite{yaf05}. Effectively, this program provides the first of several possible scenarios of minimum DCJ operations between two genomes that enables a fast and efficient process to achieve research goals. 

The representation of graphs with paths and cycles connecting synteny blocks, represented as vertices of the genomes, convey the DCJ operations to the biologist. The adjacency graph and genome graph present rearrangement scenarios to compute DCJ distance. This program expands on the data provided by Chibana \etal \cite{chibana05}, genome rearrangement of {\em C. albicans}  to {\em S. cerevisiae}. The pictorial representation of the changes in the genome graph after each operation will clarify each step for the end-user.  

Future work will include enhancing the visual effects of source and target genome genes with the use of transparency filters to accommodate visualization of DCJ operations on large data-sets. Other possible DCJ operations that fit the DCJ algorithm will also be listed, allowing the user to select their preferred path. 

\subsection*{Acknowledgment}
The authors wish to thank Rashika Mishra and Morgan Ullrich for their valuable inputs and assistance.

\bibliographystyle{abbrv}
\bibliography{sigproc}  % sigproc.bib is the name of the Bibliography in this case

\begin{thebibliography}{10}

\bibitem{Bederson04}
B.~B. Bederson, J.~Grosjean, and J.~Meyer.
\newblock Toolkit design for interactive structured graphics.
\newblock {\em IEEE Transactions on Software Engineering}, 30(8):535--546,
  2004.

\bibitem{bms06}
A.~Bergeron, J.~Mixtacki, and J.~Stoye.
\newblock A unifying view of genome rearrangements.
\newblock In {\em Proc. of the 6th International Workshop Algorithms in
  Bioinformatics, {WABI} 2006}, pages 163--173, 2006.

\bibitem{braga09}
M.~D.~V. Braga and J.~Stoye.
\newblock Counting all {DCJ} sorting scenarios.
\newblock In {\em Proc. of the International Workshop on Comparative Genomics,
  {RECOMB-CG} 2009}, pages 36--47, 2009.

\bibitem{braga10}
M.~D.~V. Braga, E.~Willing, and J.~Stoye.
\newblock Genomic distance with {DCJ} and indels.
\newblock In {\em Algorithms in Bioinformatics, 10th International Workshop,
  {WABI} 2010, Liverpool, UK, September 6-8, 2010. Proceedings}, pages 90--101,
  2010.

\bibitem{chibana05}
H.~Chibana, N.~Oka, H.~Nakayama, T.~Aoyama, B.~B. Magee, P.~T. Magee, and
  Y.~Mikami.
\newblock Sequence finishing and gene mapping for candida albicans chromosome 7
  and syntenic analysis against the saccharomyces cerevisiae genome.
\newblock {\em Genetics}, 170(4):1525--1537, 2005.

\bibitem{Compeau2013}
P.~E. Compeau.
\newblock {DCJ-I}ndel sorting revisited.
\newblock {\em Algorithms for Molecular Biology}, 8(1):6, 2013.

\bibitem{Ekdahl04}
S.~Ekdahl and E.~L.~L. Sonnhammer.
\newblock {ChromoWheel}: a new spin on eukaryotic chromosome visualization.
\newblock {\em Bioinformatics}, 20(4):576–577, 2004.

\bibitem{friedberg08}
R.~Friedberg, A.~E. Darling, and S.~Yancopoulos.
\newblock Genome rearrangement by the double cut and join operation.
\newblock {\em Bioinformatics: Data, Sequence Analysis and Evolution}, pages
  385--416, 2008.

\bibitem{UniMoG}
R.~Hilker, C.~Sickinger, C.~N. Pedersen, and J.~Stoye.
\newblock {UniMoG}—a unifying framework for genomic distance calculation and
  sorting based on {DCJ}.
\newblock {\em Bioinformatics}, 28(19):2509--2511, 2012.

\bibitem{kothari07}
M.~Kothari and B.~M. Moret.
\newblock An experimental evaluation of inversion-and transposition-based
  genomic distances through simulations.
\newblock In {\em Computational Intelligence and Bioinformatics and
  Computational Biology, 2007. CIBCB'07. IEEE Symposium on}, pages 151--158,
  2007.

\bibitem{Lin}
Q.~Lin, H.~Qi, Y.~Wu, and Y.~Yuan.
\newblock Robust orthogonal recombination system for versatile genomic elements
  rearrangement in yeast saccharomyces cerevisiae.
\newblock {\em Scientific Reports}, 5(15249), 2015.

\bibitem{Shao15}
Y.~L. Mingfu~Shao and B.~M. Moret.
\newblock {An Exact Algorithm to Compute the Double-Cut-and-Join Distance for
  Genomes with Duplicate Genes}.
\newblock {\em Journal of Computational Biology}, 22(5):425--435, 2015.

\bibitem{Overmars15}
L.~Overmars, S.~A. F.~T. van Hijum, R.~J. Siezen, and C.~Francke.
\newblock {CiVi}: circular genome visualization with unique features to analyze
  sequence elements.
\newblock {\em Bioinformatics}, 31(17):2867--2869, 2015.

\bibitem{Periwa}
V.~Periwal and V.~Scaria.
\newblock Insights into structural variations and genome rearrangements in
  prokaryotic genomes.
\newblock {\em Bioinformatics}, 31(1):1--9, 2015.

\bibitem{Sato03}
N.~Sato and S.~Ehira.
\newblock {GenoMap}, a circular genome data viewer.
\newblock {\em Bioinformatics}, 19(12):1583–1584, 2003.

\bibitem{seoighe00}
C.~Seoighe, N.~Federspiel, T.~Jones, N.~Hansen, V.~Bivolarovic, R.~Surzycki,
  R.~Tamse, C.~Komp, L.~Huizar, R.~W. Davis, S.~Scherer, E.~Tait, D.~J. Shaw,
  D.~Harris, L.~Murphy, K.~Oliver, K.~Taylor, M.-A. Rajandream, B.~G. Barrell,
  and K.~H. Wolfe.
\newblock Prevalence of small inversions in yeast gene order evolution.
\newblock {\em Proceedings of the National Academy of Sciences},
  97(26):14433--14437, 2000.

\bibitem{Shao12}
M.~Shao and Y.~Lin.
\newblock Approximating the edit distance for genomes with duplicate genes
  under {DCJ},insertion and deletion.
\newblock {\em Bioinformatics}, 13(19), 2012.

\bibitem{Stothard}
P.~Stothard and D.~S. Wishart.
\newblock Circular genome visualization and exploration using {CGView}.
\newblock {\em Bioinformatics}, 21:537--539, 2004.

\bibitem{yaf05}
S.~Yancopoulos, O.~Attie, and R.~Friedberg.
\newblock Efficient sorting of genomic permutations by translocation, inversion
  and block interchange.
\newblock {\em Bioinformatics}, 21(16):3340--3346, 2005.

\bibitem{yf09}
S.~Yancopoulos and R.~Friedberg.
\newblock {DCJ} path formulation for genome transformations which include
  insertions, deletions, and duplications.
\newblock {\em Journal of Computational Biology}, 16(10):1311--1338, 2009.

\end{thebibliography}
\end{document}